# A Neural Matrix Decomposition Recommender System Model based on the Multimodal Large Language Model

Ao Xiang

Independent researcher, Sichuan, China, xiangao1434964935@gmail.com

Bingjie Huang*

Independent researcher, Sunnyvale, CA, U.S, bingjiehuang1998@gmail.com

Xinyu Guo

Tandon School of Engineering, University of New York, Brooklyn, NY, U.S, xg897@nyu.edu

Haowei Yang

Cullen College of Engineering, University of Houston, Houston, TX, U.S, hyang38@cougarnet.uh.edu

Tianyao Zheng

Independent researcher, San Jose, CA, U.S, tian971227@gmail.com

Recommendation systems have become an important solution to information search problems. This article proposes a neural matrix factorization recommendation system model based on the multimodal large language model called BoNMF. This model combines BoBERTa's powerful capabilities in natural language processing, ViT in computer in vision, and neural matrix decomposition technology. By capturing the potential characteristics of users and items, and after interacting with a low-dimensional matrix composed of user and item IDs, the neural network outputs the results. recommend. Cold start and ablation experimental results show that the BoNMF model exhibits excellent performance on large public data sets and significantly improves the accuracy of recommendations.

## 1 INTRODUCTION

With the rapid development of the Internet, users are confronted with vast amounts of data and information. The challenge of finding content that aligns with users' interests within this abundance has become increasingly important. Recommender systems play a crucial role in addressing this issue, as they have the potential to provide precise recommendations that enhance user experience and save time in commercial applications [1]. These systems predict user ratings for specific items by employing data mining techniques and related predictive algorithms to make highly relevant predictions. By analyzing user historical behavior, preferences, and item characteristics, recommender systems effectively solve the information filtering problem by automatically matching items that may be of interest to users. Traditional recommender systems primarily consist of collaborative filtering [2], content-based recommendations [3], and hybrid recommendation methods, among which collaborative filtering is one of the earliest and most widely used techniques for recommending products or items based on past purchasing history.

However, in the era of big data, the scale of multi-modal data has become enormous, posing a greater challenge in providing accurate recommendations to users [4]. Collaborative filtering also presents limitations such as the cold start problem, sparsity, and scalability issues, which result in poor predictive performance. With the advancement of deep learning technology, more researchers are turning their focus towards neural network models to enhance the performance of recommender systems. Neural network technology has been integrated into matrix factorization, giving rise to the Neural Matrix Factorization (NMF) model [5], which has demonstrated outstanding results in recommendation prediction. However, NMF models still encounter drawbacks similar to traditional collaborative filtering methods when dealing with scenarios involving cold starts or small data volumes. As large models become more prevalent, they appear to offer a solution to the cold start problem in recommender systems. BoBERTa, a variant of the BERT large model, has exhibited strong performance across various natural language processing tasks. Therefore,this paper introduces BoBERTa into recommender systems by combining it with neural matrix factorization to propose the BoNMF model. This model harnesses the vectorization capabilities of large models effectively alleviating the cold start problem in recommender systems while enhancing recommendation accuracy and personalization.

## 2  RELATED WORK

In the field of recommender systems, matrix factorization is a classic and effective method for achieving personalized recommendations. It works by decomposing the user-item interaction matrix into low-dimensional latent feature vectors. Collaborative filtering can be divided into two main categories: user-based collaborative filtering (User-based CF), which recommends items liked by users similar to the target user, and item-based collaborative filtering, which recommends items similar to those the user likes. Matrix factorization is an implementation of collaborative filtering, with common methods including Singular Value Decomposition (SVD) and Non-Negative Matrix Factorization (NMF) [6]. The principle behind matrix factorization is to decompose the user-item rating matrix into two low-dimensional matrices to capture deep features of users and items. Collaborative filtering has always been a classic method used [7], however, traditional matrix factorization methods have been found to be difficult to deal with situations such as sparse data [8] and may not be able to effectively capture the deeper nonlinear characteristics of users and items. [9].

As neural networks have been widely used in many fields including medicine [10], business [11], intelligent driving [12], smart Contract [13], etc., recommendation systems have gradually introduced this mature technology. Neural collaborative filtering combines matrix factorization with neural networks, utilizing deep learning models to better capture complex interaction relationships between users and items. Traditional matrix factorization predicts ratings through the inner product of latent feature vectors of users and items, while neural collaborative filtering models achieve this process by constructing interaction vectors between users and items and feeding them into neural networks such as multilayer perceptron's [14]. Traditional collaborative filtering faces the cold start problem. The cold start problem refers to the situation where a recommender system cannot provide accurate recommendations for new users or new items [15]. Collaborative filtering methods rely on historical interaction data between users and items. When there is insufficient interaction data for new users or new items, collaborative filtering algorithms may fail to generate effective recommendations, leading to lower evaluation metrics for the recommender system. To address this issue, some researchers use additional information such as content and tag information to solve the cold start problem [16]. A newer approach might involve introducing large models to obtain prior knowledge [17]. Researchers have preliminarily demonstrated that large language models can effectively enhance the training signal for cold start items, significantly improving the recommendation of cold start items in various recommendation models. This study will further integrate large models with neural matrix factorization algorithms to improve the accuracy of recommender systems.ACM's new manuscript submission template aims to provide consistent styles for use across ACM publications and incorporates accessibility and metadata-extraction functionality necessary for future Digital Library endeavors. Numerous ACM and SIG-specific templates have been examined, and their unique features incorporated into this single new template. If you are new to publishing with ACM, this document is a valuable guide to the process of preparing your work for publication. If you have published with ACM before, this document provides insight and instruction into the current process for preparing` your manuscript.

## 3 METHODOLOGY

### 3.1 Dataset

To verify the effectiveness of the BoNMF model, we use an augmented dataset based on the public MovieLens dataset, which contains 1,000,209 ratings of about 3,900 movies, generated by a total of 6,040 MovieLens users. The rating information includes the movie name and category as shown in Figure 1, and the user information includes ID, gender, occupation, etc. At the same time, we added images of all movie posters and movie introductions from Wikipedia as additional supplementary information.

### 3.2 Modal Processing

BERT is a pre-trained language representation model. Unlike traditional unidirectional language models or shallow concatenation of two unidirectional language models and fixed vector encoding methods like TF-IDF, BERT employs a novel masked language model (MLM), allowing it to generate deep bidirectional language representations. The structure is shown as Figure 2. The input vector of BERT consists of token embedding, position embedding, and segment embedding. In the composition of token embedding, the first element is CLS, and the corresponding vector of the last layer of the BERT model serves as the semantic representation of the entire sentence. It can more fairly integrate the semantic information of each word in the text, better represent the emotional semantics of the entire sentence, and help the prediction of the recommendation system. RoBERTa [18] made several adjustments based on the pre-trained BERT model. These adjustments include longer training times, more training data, the removal of next predict loss, and a dynamically adjusted masking mechanism.

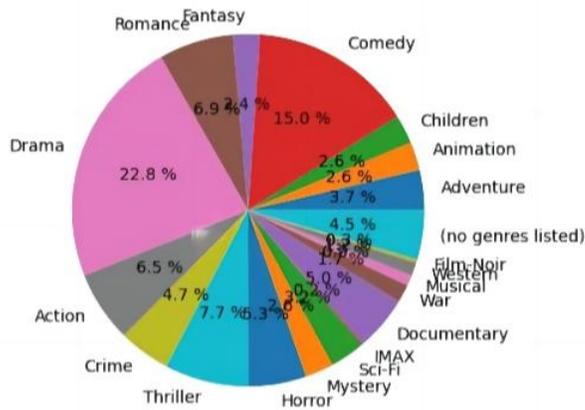

**Figure 1**. The Film Genres.

ViT[19] is also a large model with a transformer structure like BERT. It was also proposed by Google. It has a similar structure to BERT and divides the image into small blocks and then encodes the relevant positions. We use BoBERTa to extract high-dimensional feature vectors from the relevant text information of users and items and use Vit to extract relevant dimensional feature vectors from the images of items. After that, we map the user ID and item ID to a low-dimensional embedding layer, which can be regarded as the latent factors of users and items in traditional matrix decomposition. The latent factor vector of each user captures their preferences on different feature dimensions. The high-dimensional feature vector extracted from BoBERTa and the image information extracted from Vit are then fused and concatenated with the low-dimensional vector from the embedding module to form a comprehensive feature representation of users and items. These vectors are concatenated and input into the neural network to capture the interactive relationship between users and items. Finally, the fused user and item features are input into the MLP to output the user's rating prediction for the item. The structure of the BoNMF model is shown in Figure 3.

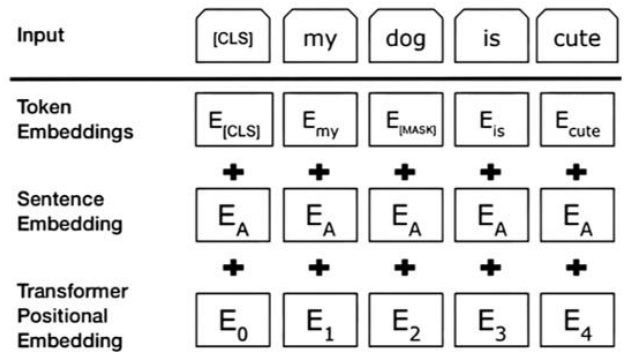

**Figure 2**. Structure of BERT embedding

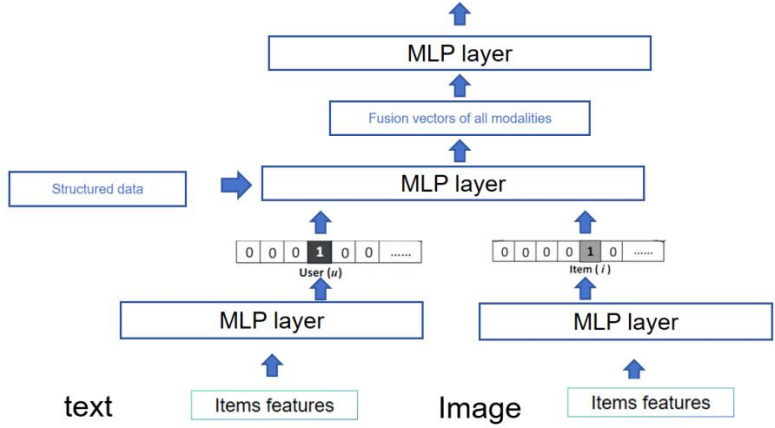

Figure 3. The structure of BoNMF model

### 3.3 Evaluation Metrics

In the task, three methods commonly used evaluation metrics, mean square error, average accuracy, and precision@K, are applied in this paper.

Mean squared error (MSE), which shown in Formula 1, a metric used to evaluate the performance of a regression model and measures the average squared difference between predicted and actual values. It can reflect the overall prediction accuracy of the model and is the most widespread evaluation index. $n$ is the total number of samples, $\hat{r}_i$ is the predicted result of the i-th observation predicted by the model, and $r_i$ is the actual value of the i-th observation.

$$\text{MSE} = \frac{1}{n}\sum_{i=1}^{n}(\hat{r}_i - r_i)^2 \quad \#(1)$$

Precision@K is an indicator used to evaluate the performance of recommendation systems, measuring the proportion of the number of relevant items in the top K recommendation results to the total number of recommendations. U represents the set of users, $R_u$ is the interaction between the user and the items, and $R_u \cap \widehat{R_{u,K}}$ is the results of recommended items that the user is interested in $R_u$ is the interaction matrix between the user and the item, and $R_u \cap \widehat{R_{u,K}}$ is the final recommended project results with the highest ranking.
.

$$\text{Precision@K} = \frac{1}{|U|}\sum_{u \in U}\frac{|R_u \cap \widehat{R_{u,K}}|}{K} \quad \#(2)$$

Normalized Discounted Cumulative Gain (NDCG) is also popular metric for evaluating the quality of recommendation systems that takes into account the relevance and position of the results. DCG@K (Discounted Cumulative Gain) is used to calculate the cumulative gain of the top k recommended items for a given user. IDCG@K (Ideal Discounted Cumulative Gain) represents the maximum possible value of DCG@K under ideal conditions. By assigning weights to positions, the highest-ranked correlations receive greater importance, reflecting the significance of more relevant rankings. Among them, DCG calculates the sum of correlations in the recommended results, regardless of the location of the results. DCG introduces the total correlation in the recommendation results of the location loss factor.

$$\text{NDCG@K} = \frac{1}{|U|} \sum_{u \in U} \frac{\text{DCG@K}}{\text{IDCG@K}} \#(3)$$

## 4 EXPERIMENT

The user's merged embedding vector and the movie's BoBERTa embedding vector are both 769-dimensional, and the user's movie id is reorganized into a 50-dimensional embedding vector through MLP. The two are directly spliced and then output through a 128-dimensional MLP. ll MLPs were configured with a two-layer structure: the first layer consisting of 128 neurons and the second layer consisting of 64 neurons. The third layer served as the output layer. We split the dataset into training and testing sets in a 7:3 ratio and employed cross-validation for training. and the number of recommended items, was set to 5, all neural networks were trained for 10 epochs. At the same time, the study removed part of the movie's modal information for comparison, removing images, text, and structural data respectively. At the same time, to evaluate the effectiveness of the proposed model, we conducted multiple ablation experiments to test the importance of different modalities on model performance and the impact of adding LLM on improving semantic vector representation.

First, we directly used traditional matrix decomposition as the baseline model. At the same time, we also added a simple neural matrix decomposition model as another comparison. Simple neural matrix decomposition the vector representation of large model decomposition will not be used but will be directly converted into low-dimensional vectors through MLP and then spliced. We also added three models that removed images, text, and structured data. Finally, we compared some recently popular methods for splicing different large models, such as Tensor product-based model [20], Text Early Fusion Model [21] and DNN - based model [22]. The Tensor product-based model uses a large language model to combine multiple modalities in the form of vector products. The Text Early Fusion Model uses an early fusion approach to concatenate text and low-dimensional vectors for training, while the DNN-based model directly maps text into vectors and directly enters the deep network for training after performing dot products. The final model training results are shown in the table 1.

Table 1. Ablation experiment results

| Model | MSE | Precision@K | NDCG |
| --- | --- | --- | --- |
| BoNMF | 0.6917 | 0.8865 | 0.6911 |
| NMD | 0.8075 | 0.8355 | 0.6367 |
| SVD | 5.011 | 0.8465 | 0.6184 |
| Tensor product-based | 0.7212 | 0.8510 | 0.6559 |
| Text Early Fusion | 0.7212 | 0.8418 | 0.6559 |
| CNN-based | 0.9421 | 0.8210 | 0.6224 |
| BoNMF -no Figure | 0.7112 | 0.8677 | 0.6331 |
| BoNMF-no Text | 0.7258 | 0.8512 | 0.6224 |

## 5 ANALYSIS

The BoNMF model exhibits the MSE of 0.6917, while the NMD model shows an MSE of 0.8075. The SVD model has the highest MSE at 5.011, significantly surpassing the other two models. This indicates that the BoNMF model achieves the smallest average error between its predicted ratings and the actual ratings, demonstrating superior accuracy in capturing user rating behaviors. The NMD model, which employs a neural network structure, substantially alleviates the shortcomings associated with the absence of high-dimensional vector representations. In contrast, the traditional SVD model fails to capture deeper semantic meanings effectively.

Additionally, the BoNMF model attains the highest NDCG@K of 0.6911, indicating that its recommendation results are the most relevant and well-ranked. The NMD model achieves an NDCG@K of 0.6367, slightly lower than that of the BoNMF model, while the SVD model's NDCG@K is 0.6184, placing it between the BoNMF and NMD models, indicating relatively good relevance and ranking of its recommendation results. Lastly, the Precision@K for the BoNMF, NMD, and SVD models is uniformly 0.8865, indicating that all three models have the same accuracy in the top K recommendations. However, this does not imply that the effectiveness of the three models is identical. This uniformity could be attributed to the dataset's lack of long sentence semantic information or the evaluation function's leniency, which does not account for the weight of different factors. Therefore, future experiments should aim to improve the completeness of the experimental dataset by including more subjective and abstract semantic information to enhance the recommendation system's accuracy.

When the BoNMF model does not include image information, the MSE increases to 0.7112, and Precision@K and NDCG decrease to 0.8677 and 0.6331, respectively. Although the precision and NDCG decrease, the difference is not particularly large compared to the BoNMF model that includes image information. When the BoNMF model does not include text information, the MSE increases to 0.7258, and Precision@K and NDCG decrease to 0.8512 and 0.6224, respectively. Compared with the full BoNMF model, the performance decreases significantly, which indicates that the description is more accurate than the information presented in the image. Secondly, the addition of the image does not significantly improve the accuracy and recommendation level, so it is reasonable to guess that most of the elements that appear in the image may already be included in the description.

## 6 DISCUSSION

We can see that compared with some popular fusion methods currently in use, our model still has certain advantages. Although not shown in the table, we also compared the impact of different large models such as BERT and Xlnet on high-dimensional vector decomposition with some mainstream large models, but according to the results, the difference is within 0.5%, which may still be caused by defects such as data sets; It must be admitted that for different modes, some models such as Swin-T model will make the progress higher and higher, and the impact of large models with different structures on high-dimensional vector decomposition needs further discussion.

In summary, the BoNMF model performs excellently across various metrics, making it the optimal choice and proving its superiority. It is noteworthy that the SVD model's NDCG is higher than that of the NMD model. This may suggest that replacing matrix factorization with a neural network model might lead to a decrease in the NDCG metric, the specific reasons for which require further investigation.

In the experiment, we also found that although the text-based movie introductions can already achieve good performance, the addition of image information such as posters can still improve on different models. But the improvement is minimal. An interesting discovery is that on the poster, there is actually information including the starring information, time, and movie name, which will conflict with the low-density low-bit vector information to a certain extent, resulting in poor performance. Compared with last year, the current emphasis on multimodal images needs to be improved. In future work, more consideration should be given to extracting the real local features of the image, rather than including parts that conflict with text information. This may require further research on image positioning.[23] Meanwhile, we plan to further optimize algorithms related to visual aspects, specifically enhancing ViT's semantic understanding of different regions to better comprehend the contribution of various modalities to the accuracy of the recommendation system from a deeper perspective [24]. In addition, we find that information visualization between different modules is also a direction worth exploring. It can reveal more information overlap problems between different modalities to improve the prediction accuracy of the model, and at the same time it can make the model more concise and we believe that persisting in solving the problem of modal overlap can better make greater contributions in some scenarios such as application contract execution [25] and provides better support and visual response for some algorithms that need to call multimodal data [26]..

## 7  CONCLUSION

In this study, we introduce the BoNMF model, a neural matrix factorization recommendation system based on large-scale pre-trained models BoBERTa and ViT. By integrating the high-dimensional text feature vectors from BoBERTa and the high-dimensional image feature vectors decomposed by ViT and combining them with the low-dimensional embeddings of user and item IDs, this model effectively leverages the capabilities of large models in understanding text and images and captures deep feature representations of users and items through the advantages of neural matrix factorization networks. Our experimental results on the MovieLens dataset demonstrate that the BoNMF model outperforms the NMD and SVD models across various evaluation metrics, achieving an impressive MSE of 0.69. Our findings indicate that the BoNMF model is a promising approach for improving recommendation systems, especially in addressing the cold-start problem and capturing deep semantic information. Future work will focus on further exploring the utilization of multimodal data to provide deeper insights for optimizing recommendation systems